\documentclass{elsart}
\usepackage{graphicx}
\usepackage{amsmath}
\usepackage{color}
\usepackage[dvipsnames]{xcolor}
\usepackage{bm}

\usepackage{cite}

\newcommand{\rmi}{{\rm i}}

\newcommand{\zr}{z_{\text{\tiny R}}}
\newcommand{\gr}{\Gamma _{\text{\tiny R}}}

\newcommand{\tr}{\tau _{\text{\tiny R}}}
\newcommand{\er}{E_{\text{\tiny R}}}

\newcommand{\vws}{V_{\text{\tiny WS}}}
\newcommand{\vso}{V_{\text{\tiny SO}}}

\journal{Nuclear Physics A}

\begin{document}
\begin{frontmatter}
\title{The Gamow-state description of the decay energy spectrum of 
neutron-unbound $^{25}$O}

\author{R.M.~Id Betan,$^{1,2,3}$ R.~de la Madrid$^{4}$}
\address{$^{1}$Physics Institute of Rosario (CONICET), 
             	Esmeralda y Ocampo, S2000EZP Rosario, Argentina.}
\address{$^{2}$Department of Physics FCEIA (UNR),
             	Av.~Pellegrini 250, S2000BTP Rosario, Argentina.}
\address{$^{3}$Institute of Nuclear Studies and Ionizing Radiations (UNR), 
		    	Riobamba y Berutti, S2000EKA Rosario, Argentina.}
\address{$^{4}$Department of Physics, Lamar University, Beaumont, TX 77710,
United States.}

\date{\today}

\begin{abstract}
We show the feasibility of calculating the decay energy spectrum of 
neutron emitting nuclei within the Gamow-state description of
resonances by obtaining the decay energy spectrum of $^{25}$O. We
model this nucleus as a valence neutron interacting with an 
$^{24}$O inert core, and we obtain the resulting
resonant energies, widths and decay energy spectra for the ground and 
first excited states. We also discuss the similarities and differences between
the decay energy spectrum of a Gamow state and
the Breit-Wigner distribution with energy-dependent width. 
\end{abstract}

  
\end{frontmatter}

\section{Introduction} 
\label{sec.introduction}

Ever since the discovery of the exotic $^{11}$Li nucleus~\cite{1986Tanihata},
there have been many experimental~\cite{2003Jonson,BAUMANN12,TANIHATA13} and
theoretical~\cite{2003Okolowicz,2014Volya,MICHEL09} studies of nuclei far
from the beta stability line. On the experimental side, new techniques were 
developed to produce and study the properties of rare isotopes. On the 
theoretical side, new models were developed to 
guide and explain the experimental findings. The theoretical and experimental 
understanding of unstable nuclei will continue to be one of the main goals 
of the nuclear physics community~\cite{2013Nap}, as testified by the 
construction and updating of a number of facilities around the 
world~\cite{2000Mueller,2000Nupecc,2013Blumenfeld,2014Motobayashi}.

Experimentally, one method of studying unstable nuclei is by means of 
their decay energy spectrum, which measures the number of decays per unit of 
energy versus energy (see for example Refs.~\cite{PEREZ16,VANDEBROUCK17}). In
particular, by using techniques such as invariant-mass spectroscopy, it
is possible to obtain the experimental decay energy spectrum  of some
unstable nuclei that decay by neutron emission (see for example
Refs.~\cite{BAUMANN12,SPYROU10,SPYROU12,THOENNESSEN12}). Hence,
it is important to be able to calculate such spectra theoretically. 

In Refs.~\cite{NPA15,NPA17}, the resonant (Gamow) state was
used to obtain a theoretical expression for the decay energy spectrum of an
unstable system decaying into the continuum. The purpose of the present paper
is to use the formalism
of Refs.~\cite{NPA15,NPA17} to obtain the energy spectrum of
the $^{25}$O nucleus, which decays by neutron emission.

Dripline Oxygen isotopes are currently of great interest, both
theoretically and experimentally. From the excited states of
$^{19}$O~\cite{2016Dungan} to the 
isotopes well beyond the neutron drip line~\cite{2017Fossez}, there are several
Oxygen isotopes whose energies, widths, and decay energy spectra can be used
as a test bench for different theories. The heavier neutron drip line nucleus
that has been observed experimentally is $^{24}$O, which was found to be 
doubly magic~\cite{2009Hoffman}. An excited state of $^{24}$O was 
found~\cite{JONES15} to decay sequentially to $^{22}$O. In addition, 
other low-lying neutron-unbound excited states of $^{24}$O have been 
measured~\cite{ROGERS15}. The energy and width of the unbound ground state 
of $^{25}$O were investigated in
Refs.~\cite{HOFFMAN08,CAESAR13,KONDO16,2017Jones}, and strong evidence for
the first excited state of $^{25}$O was found in Ref.~\cite{2017Jones}. The
ground state~\cite{LUNDERBERG12,KONDO16}, excited states and decay
modes~\cite{THOENNESSEN12,CAESAR13} of $^{26}$O have also been studied. 

In our analysis, we will use the neutron-unbound $^{25}$O because its decay
energy spectrum has been measured 
experimentally~\cite{HOFFMAN08,CAESAR13,KONDO16,2017Jones}. Since $^{24}$O is 
doubly magic, we will treat $^{25}$O as a valence neutron in an $^{24}$O 
core, and we will describe the valence neutron by a Gamow state. This 
two-body model is able to reproduce the experimental
ground state of $^{25}$O, 3/2$^+$. However, our two-body model yields 7/2$^-$ 
as the first excited state, instead of the one found experimentally, 1/2$^+$.

The structure of the paper is as follows. In Sec.~\ref{sec.formalism}, we 
summarize the formalism needed to calculate the theoretical decay energy 
spectrum. In Sec.~\ref{sec.validation},
we assess the validity of the code. In Sec.~\ref{sec.oxygen25}, we apply
the code to the $^{25}$O nucleus. We will model the interaction between the
valence neutron and the $^{24}$O core by the Woods-Saxon and the spin-orbit
potentials, and obtain the energies, widths and decay energy spectra of the
ground and first excited states. We will compare our results with
those obtained by the Continuum and the Gamow Shell 
Models~\cite{MICHEL09,2014Volya}. We will also compare the 
theoretical decay energy spectrum with the experimental 
ones~\cite{HOFFMAN08,CAESAR13,KONDO16,2017Jones}. In addition, we will
discuss the similarities and differences (both quantitative and 
phenomenological) between the Gamow-state decay energy spectra and
the Breit-Wigner distributions with energy-dependent width. In
Sec.~\ref{sec.conclusions}, we summarize our main results and present 
an outlook of future applications.


\section{Formalism} 
\label{sec.formalism}

In order to be self-contained, in this section we outline the main 
ingredients needed to calculate the decay energy
spectrum of $^{25}$O.

\subsection{The decay energy spectrum of a Gamow state}

Let $H=H_0 + V$ be the Hamiltonian that describes the decaying system, formed
in our model by an inert $^{24}$O core and a valence neutron. The free 
Hamiltonian $H_0$ is the part of the Hamiltonian that describes the 
valence neutron after it has been ejected and is far away from the core,
whereas $V$ is the interaction potential between the core and the valence
neutron. Let us describe the unstable state by a Gamow~\cite{1928Gamow}
state $|\zr \rangle$ such that
$H |\zr \rangle = \zr |\zr \rangle$ and $\zr =\er - \rmi\, \gr /2$. Then,
the differential decay width $\frac{d\overline{\Gamma}}{dE}$, which
describes the strength of the interaction between the resonance and
the continuum at each scattering energy $E$, is given by~\cite{NPA15}
\begin{equation}
    \frac{d\overline{\Gamma}(E)}{dE} = 2\pi L(E) 
              | \langle E | V | \zr \rangle |^2 \, , 
       \label{eq.ddw}
\end{equation}
where $L(E)$ is the Lorentzian distribution
\begin{equation}
   L(E) = \frac{1}{\pi} 
             \frac{\gr /2}{(E - \er )^2 + (\gr /2)^2} \, , 
\end{equation}
and $|E\rangle$ is an eigenstate of the free Hamiltonian with energy
$E$, $H_0|E\rangle =E|E\rangle$.

The normalized, theoretical decay energy spectrum of a resonance decaying 
into the continuum is then obtained as~\cite{NPA15}
\begin{equation}
  \frac{dP(E)}{dE} = \frac{1}{\overline{\Gamma}}
                       \frac{d\overline{\Gamma}(E)}{dE} 
    = \frac{2\pi}{\overline{\Gamma}} L(E) \, 
             | \langle E | V | \zr \rangle |^2 \, , 
      \label{eq.des}
\end{equation}
where $\overline{\Gamma}$ is the total decay width,
\begin{equation}
     \overline{\Gamma} = \int_0^\infty 
                   \left(  \frac{d\overline{\Gamma}}{dE}  \right) dE 
     = 2\pi \int_0^\infty
             L(E)  | \langle E | V |\zr \rangle |^2 dE \, .
    \label{eq.tdw}
\end{equation}
We will identify the theoretical spectrum of Eq.~(\ref{eq.des}) with the
experimental decay energy spectrum.

It should be noted that $\overline{\Gamma}$ is in general different from 
$\gr$, and therefore $\overline{\Gamma}$ is not related to the lifetime
of the resonance. Physically, we can interpret $\overline{\Gamma}$ as a
measure of the overall strength of the interaction between the resonance
and the continuum. This is why in Sec.~\ref{sec.oxygen25} we will use 
$\overline{\Gamma}$ to quantify
the relative strength of the spectra of different resonances.

It should also be noted that Eqs.~(\ref{eq.des}) and~(\ref{eq.tdw}) represent,
in a way, an extension of Fermi's Golden Rule to the case where the 
broadening of the decay energy spectrum is taken into account. Such 
broadening is provided by the Breit-Wigner distribution.

\subsection{The Schr\"odinger equation of the valence neutron}

Because $^{24}$O is a doubly magic nucleus~\cite{2009Hoffman}, we are going 
to neglect the many-body nature of $^{25}$O and treat it as an $^{24}$O 
core plus a single neutron that is subject to a mean-field potential 
created by the $^{24}$O core. As it is customary, we will model such mean-field 
potential by the Woods-Saxon potential and a spin-orbit interaction,
\begin{equation}
   V(r)= \vws (r) + \vso (r)=  -V_0 f(r) + V_{\rm so} \frac{1}{r} \frac{df(r)}{dr} \xi_{l,j}
   \, ,
   \label{eq.se}
\end{equation}
where $V_0>0$ represents the potential well depth, and $V_{\rm so}>0$ 
represents the strength of the spin-orbit interaction. The function
$f(r)$ is given by 
\begin{equation}
    f(r) = \frac{1}{1+{\rm exp}\left( \frac{r-R}{a}\right)} \, ,
    \label{fofr} 
\end{equation} 
where $a$ is the diffuseness parameter (or surface thickness), and 
$R=r_0A^{1/3}$ is the nuclear radius, $A$ being the mass number. The
function $\xi_{l,j}$ is given by
\begin{equation}
  \xi_{l,j}=
  \left\{
  \begin{array}{ccc}
   \frac{l}{2} & \quad  \textnormal{for} & j=l+\frac{1}{2}  \, ,  \\
    -\frac{(l+1)}{2} & \quad \textnormal{for} & j=l-\frac{1}{2}  \, , 
  \end{array}
  \right.
\end{equation}
where $l$ and $j$ are the orbital and the total angular momentum of
the valence neutron, respectively.

Due to the spherical symmetry of the potential, we can work with 
spherical coordinates, separate the radial and angular dependences, and 
obtain the radial Schr\"odinger  equation for each partial wave,
\begin{equation}
\biggl(\frac{-\hbar^2}{2\mu}\frac{d^2}{dr^2}+\frac{\hbar^2l(l+1)}{2\mu r^2}
	+V(r) \biggr) u_l(r;E)=E u_l(r;E)\, ,
	\label{baba}
\end{equation}
where $\mu$ is the reduced mass of the system. By solving Eq.~(\ref{baba}) 
subject to purely outgoing boundary  conditions, we obtain the resonant 
(Gamow) eigenfunctions $u_l(r;\zr)$. When $\zr$ is real and negative, 
$u_l(r;\zr)$ becomes a bound state. Under the appropriate boundary conditions,
Eq.~(\ref{baba}) yields the scattering eigenfunctions when $E$ is positive.

After the neutron is expelled from the nucleus, it behaves like a free 
particle, and therefore its radial wave function $\chi_l(r;E)$ 
satisfies the radial, free Schr\"odinger equation,
\begin{equation}
     \biggl(\frac{-\hbar^2}{2\mu}\frac{d^2}{dr^2}+\frac{\hbar^2l(l+1)}{2\mu r^2}
       \biggr) \chi_l(r;E)=E \chi_l(r;E) \, ,
  \label{schroeqfree}
\end{equation}
subject to the boundary condition that the eigenfunction is regular at the 
origin, $\chi_l(0;E)=0$. The delta-normalized solution of 
Eq.~(\ref{schroeqfree}) that is regular at the origin is given by the 
reduced Riccati-Bessel function $\hat{j}_l$ (see for example 
Ref.~\cite{TAYLOR}),
\begin{equation}
  \chi_l(r;E)= \sqrt{\frac{2\mu}{\hbar ^2}} \frac{1}{\sqrt{k\pi}}
                 \hat{j}_l (kr) \, ,
          \label{freeeign}
\end{equation}
where $k=\sqrt{\frac{2\mu}{\hbar ^2}E}$ is the wave number. The 
Riccati-Bessel function can be written~\cite{TAYLOR} in terms of the 
spherical Bessel function $j_l(z)$ and the ordinary Bessel function 
$J_{\lambda}(z)$ as $\hat{j}_l (z)= zj_l(z)= \sqrt{\frac{\pi z}{2}} J_{l+1/2}(z)$.

By combining Eq.~(\ref{eq.des}) with the Gamow eigenfunction $u_l(r;\zr)$ and
with the free, radial eigenfunction $\chi_l(r;E)$, we can obtain the theoretical
decay energy spectrum of $^{25}$O.

\section{Validation of the numerical procedure}
\label{sec.validation}

We have used the code {\sc Gamow}~\cite{1982Vertse} to solve numerically the 
Schr\"odinger equation (\ref{baba}) in order to obtain the resonant 
energies and the Gamow states. We have used the code 
{\sc Anti}~\cite{1995Ixaru,1996Liotta} to obtain the scattering states. The 
resulting energies and eigenfunctions were afterward plugged into 
Eq.~(\ref{eq.des}) to obtain the numerical decay energy spectrum. Because the
Gamow eigenfunctions diverge exponentially, and because the resonant 
energies are usually very sensitive to small changes in the parameters of the
potential, we performed three tests to validate our numerical procedure.

\subsection{First test: The energy density of the free eigenfunctions}

\begin{figure}[h!]
\vspace{6mm}
\begin{center}
        \includegraphics[angle=0,width=0.65 \columnwidth]{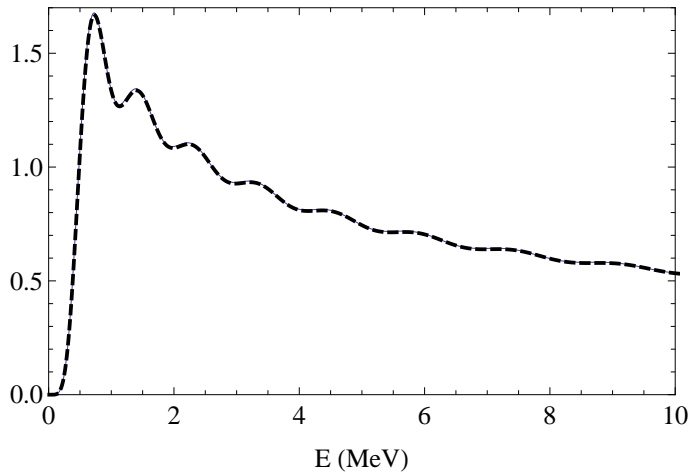}
\end{center}                
\caption{(Color online) Comparison of the integral 
$\int_0^{50} |\chi_l(r;E)|^2  \, dr$ for $l=6$ and 
$\frac{2\mu}{\hbar^2}=0.047892$ MeV$^{-1}$-fm$^{-2}$ obtained  numerically 
using Mathematica (thin, blue line)
and the code {\sc Anti} (thick, dashed, black line). The plots are essentially
indistinguishable.}
\label{fig:freedensitymath}
\vspace{6mm}
\end{figure}

In the first test, we calculated the energy density of the free 
scattering eigenfunctions, $\int dr \,  |\chi_l(r;E)|^2$, using 
the code {\sc Anti}~\cite{1995Ixaru,1996Liotta} and
Mathematica~\cite{Mathematica}. We have plotted the 
results for $l=6$ in Fig.~\ref{fig:freedensitymath},
where we can see that the plots are essentially the same. In particular,
{\sc Anti} and Mathematica yield a maximum of 0.7254 MeV$^{-1}$ at the 
energies of 1.677~MeV and 1.65~MeV, respectively. 

\subsection{Second test: Location of the resonant energy in the 
energy density of the scattering eigenfunctions}

\begin{figure}[h!]
\begin{center}
        \includegraphics[angle=-90,width=0.65\columnwidth]{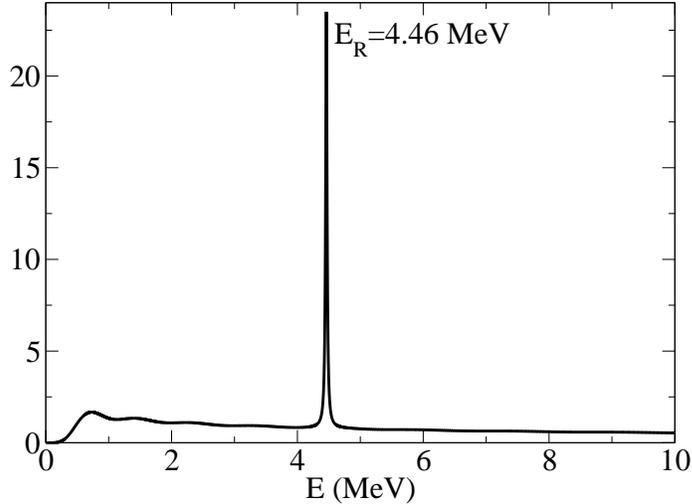}
\caption{Plot of the energy density of the 
scattering eigenfunction, $\int_0^{50} | u_l(r;E)|^2 \, dr $.}
\label{fig:intescattersq}
\end{center}                
\end{figure}

In order to test the accuracy of the resonant energies, we compared the real 
part of such energies with the peaks in the energy density of 
the scattering eigenfunctions. In this second test, we fixed the parameters
of the Woods-Saxon potential so that we can reproduce the 
lowest energy levels of $^{133}$Sn. For simplicity, in this second test 
we neglected the spin-orbit interaction. The values of the parameters
we used are $V_0= 43.5$~MeV and $R=6.466$~fm ($r_0=1.27$ fm). For $l=6$, 
we obtained a sharp resonance of complex energy 
$\zr = (4.460- \rmi \, 0.014)$ MeV. We then obtained the scattering 
eigenfunctions $u_l(r;E)$ of Eq.~(\ref{baba}) for $l=6$. The 
resulting energy density, $\int | u_l(r;E)|^2 \, dr $, where $r$ is 
in fm and $E>0$ is in MeV, is plotted in Fig.~\ref{fig:intescattersq}.

It is clear from Fig.~\ref{fig:intescattersq} that the radial probability
density of the scattering wave function is sharply peaked around the 
energy $4.46$~MeV. This energy coincides with the real part of the 
resonant energy $\zr = (4.460- \rmi \, 0.014)$ MeV.

\subsection{Third test: The delta-shell potential}
\label{subsec.third}

The formalism of Ref.~\cite{NPA15} was applied in Ref.~\cite{NPA17}
to the delta-shell potential $V_\delta(r) = g \, \delta(r-R)$ for $l=0$. It was
found in Ref.~\cite{NPA17} that the $s$-wave resonant energies of the 
delta-shell
potential can be expressed in terms of the Lambert $W$ function, and therefore
one can calculate the resonant energies, decay widths and decay energy spectra 
of $V_{\delta}(r)$ exactly. As a third validation 
of our numerical procedure, we have applied it to an almost delta-shell 
potential\footnote{In Appendix~\ref{sec.appendix}, we explain in what 
sense the potential of Eq.~(\ref{almostdelta}) 
is almost a delta-shell potential.} and compared the 
results with those of Ref.~\cite{NPA17}. 

Our almost delta-shell potential is given by
\begin{equation}
    V_a(r) = - g \frac{d f(r)}{d r}= 
            \frac{g}{a} 
\frac{e ^{ \frac{r-R}{a}}}{\left[ 1+e ^{\frac{r-R}{a}} \right] ^2} \, , 
          \label{almostdelta}
\end{equation}
where $a$ is very small. It should be noted
that the choice $g>0$ ($g<0$) makes the potential repulsive (attractive). As 
explained in Appendix~\ref{sec.appendix}, when 
$a$ is very small, $V_a(r)$ becomes, for practical purposes, the 
delta-shell potential centered at $r=R$.

Since the potential $V_a(r)$ becomes very singular when $a$ is small,
care must be taken in obtaining the resonant energies when $a$ 
tends to zero. In our case, we used a two-step process to obtain the 
resonant energies. In the first step, we obtained the resonant 
energies for decreasing values of $a$; in the second step, we 
extrapolated~\cite{nr} the 
result to $a \rightarrow 0$. In this way, we first calculated the $l=0$ ground
(bound) and first excited (unbound) states for the case that 
$R=6.466$~fm and $\frac{2\mu}{\hbar^2}=0.047892$~MeV$^{-1}$-fm$^{-2}$ 
(which correspond to $^{133}$Sn), and for decreasing values of $a$ 
up to $a=0.04$~fm. Afterward, we extrapolated the results using four order 
algebraic extrapolation~\cite{nr} up to $a=10^{-5}$~fm. 

In order to compare our results with those of Ref.~\cite{NPA17}, we need to 
recall that the results of Ref.~\cite{NPA17} were given in terms of the 
dimensionless coupling constant $\lambda=\frac{2\mu}{\hbar^2}Rg$, Hence, 
with our choice of $R$ and $\mu$ ($R=6.466$~fm and 
$\frac{2\mu}{\hbar^2}=0.047892$~MeV$^{-1}$-fm$^{-2}$), 
the strength $g$ can be written in terms of $\lambda$ as 
$g=3.229\, \lambda$~MeV-fm, and the energies and widths of 
Ref.~\cite{NPA17} are given in units of $\hbar^2/2\mu R^2=0.4994$~MeV.

In Table~\ref{table.energies}, we compare the calculated ground state 
and first-excited state energies with those of Ref.~\cite{NPA17} for
$\lambda=-0.5,\, -10,$ and $-100$. As can 
be seen in Table~\ref{table.energies}, our numerical results are in
fairly good agreement with the exact ones.

\vskip0.5cm

\begin{table}[ht] 
\begin{center}
\caption{Comparison of the calculated ground state energy $E_{\rm gs}$ 
and first excited state energy $\zr$ with those of Ref.~\cite{NPA17}.}

\begin{tabular}{|cc|cc|cc|}
\hline
 \multicolumn{2}{|c|}{Strength}     & \multicolumn{2}{c|}{$E_{\rm gs}$ (MeV)}  & 
        \multicolumn{2}{c|}{$\zr$ (MeV)}    \\
    $\lambda$ & $g$ (MeV) &  Exact                    & Present work   
                      &  Exact                    & Present work \\
\hline
    $-$0.5    & $-$1.6146     &  $-$0.19711       &  $-$0.20554  
   &  5.935 $-$ i 5.180 & 5.895  $-$ i 5.204 \\
     $-$10           & $-$32.29           &  $-$12.484     & $-$12.475  &
     												5.897 $-$ i 0.356 & 5.823   $-$ i 0.335 \\
     $-$100         & $-$322.9           &  $-$1248.5      &  $-$1139.0 &
     											   5.029 $-$ i 0.00320 & 5.030 $-$ i 0.00298\\
\hline
\end{tabular}
\end{center}
\label{table.energies}
\end{table}

\vskip1cm

The calculation of the decay width of Eq.~(\ref{eq.tdw}) involves 
the resonant states and the free scattering states, and therefore 
constitutes a more demanding test. In Table~\ref{table.tdw}, we compare 
the calculated decay width with the exact one for the first excited state
when $\lambda=-0.5,\, -10,$ and $-100$.

\begin{table}[ht]
\caption{Comparison of the calculated $\overline{\Gamma}$ with that of 
Ref.~\cite{NPA17} for the first excited state.} 
\begin{center}
\begin{tabular}{|cc|cc|}
\hline
   \multicolumn{2}{|c|}{Strength}    &    \multicolumn{2}{c|}{$\overline{\Gamma}$ (MeV)}  \\
    $\lambda$ & $g$ (MeV)  &  Exact      &  Present work \\
\hline
    $-$0.5  & $-$1.6146          &  0.22769    & 0.0688 \\
     $-$10  &  $-$32.29       & 0.9335    & 0.825 \\
     $-$100 &  $-$322.9     &  0.01283     &  0.0127\\
\hline
\end{tabular}
\end{center}
\label{table.tdw}
\end{table}

As can be seen in Table~\ref{table.tdw}, our calculated decay width
agrees well with the exact one for strong couplings ($\lambda = -10, \, -100$)
but not for weak couplings ($\lambda = -0.5$). The reason 
why our calculated $\overline{\Gamma}$ is not accurate when $\lambda$
is small is that, for weak couplings, the resonance is broad, its
decay energy spectrum is also broad, and hence its tails are not negligible
at energies much higher than the resonant energy. Since our numerical procedure
to calculate $\overline{\Gamma}$ omits the high-energy tails, our 
decay constant is much smaller than the exact one when the coupling
is weak.

Overall, our numerical results are accurate for sharp resonances, but not for
broad ones. However, since the resonances of $^{25}$\rm{O} 
are sharp, we expect that our numerical results for $^{25}$\rm{O} to
be fairly accurate, except for underestimating the decay widths 
$\overline{\Gamma}$ due to 
the neglect of the high-energy tails of the decay energy spectrum.

\section{The decay energy spectrum of $^{25}$\rm{O}}
\label{sec.oxygen25}

In this section, we are going to calculate the decay energy spectrum of the 
ground state and of the first excited state of the unbound oxygen isotope 
$^{25}$O using a simple two-body model, i.e., one valence neutron outside the 
$^{24}$O core that creates the potential of
Eq.~(\ref{eq.se}). 

We choose the radius and diffuseness of the mean field potential
(the same for the Woods-Saxon and the spin-orbit parts) as in
Ref.~\cite{2006Volya}, $a=0.65$ fm and $r_0=1.06$ fm. The Woods-Saxon
$V_0$ and spin-orbit $V_{\rm so}$ strengths are chosen to approximately
account for the average $768.5$ keV of the experimental ground state energies
of $^{25}$O reported in Refs.~\cite{HOFFMAN08} ($770$ keV),
\cite{CAESAR13} ($725$ keV), \cite{KONDO16} ($749$ keV) and \cite{2017Jones}
($830$ keV), and for the experimental gap between the ground state energy
of $^{25}$O with that of the first hole state in the $^{24}$O core,
$E_{0d_{3/2}} - E_{1s_{1/2}}=4.857$~MeV~\cite{HOFFMAN08}. Such criteria, and
the experimental neutron separation energy
$S_n(^{25}\rm{O})=-0.776$ MeV~\cite{nndc}, lead to the following parameters:
$V_0=57.7$~MeV, and $V_{\rm so}=15.32$~MeV-fm$^2$.

For the above parameters, the complex energy of the ground state was found 
to be $z_{d_{3/2}}=(0.766 - \rm{i}\, 0.034)$~MeV. Thus, the ground state's pole
width is $\Gamma _{\text{\tiny R},d_{3/2}} = 68$~keV. This pole width is similar 
to the pole width of the Continuum Shell Model~\cite{2014Volya} 
($63$ keV), and slightly higher than that of the Gamow Shell Model (the
average of the two models used in Ref.~\cite{2017Fossez} yields $49.5$ keV,
whereas Ref.~\cite{2017Jones} reports a pole width of 
$51$~keV). 

Comparison of $\Gamma _{\text{\tiny R},d_{3/2}}$ with the experimental pole widths
is not so straightforward, because the experimental pole widths vary wildly 
($172$ keV in Ref.~\cite{HOFFMAN08}, $20$ keV in Ref.~\cite{CAESAR13}, and 
$88$ keV in Ref.~\cite{KONDO16}). Their average, 93~keV, is larger than
any theoretical pole widths. In fact, the experimental pole 
widths are usually overestimated, because the experimental error is 
convoluted with the ideal decay energy spectrum, which makes the experimental
decay energy spectrum broader than the ideal one.

In the four-body model of Ref.~\cite{2017Jones}, the first excited state of 
$^{25}$O was reported to be a $1/2^+$ state, and strong experimental
evidence for such state was also found~\cite{2017Jones}. However,
our simple two-body model is unable to produce such 
$1/2^+$ as the first excited state.\footnote{The $1/2^+$ state may be seen
as an excitation of the $^{24}$O core~\cite{1991Frauenfelder}, which in our 
model is inert.} Instead, the first excited state
of our two-body model is $f_{7/2}=7/2^-$, whose complex energy is 
$z_{f_{7/2}}=(5.588 - \rm{i}\, 0.697)$ MeV. The four-body model of
Ref.~\cite{2017Jones} also produces an $f_{7/2}=7/2^-$ state, whose complex
energy is $5.536 - \rm{i}\, 0.0075$ MeV. Thus, the energy
predicted by our two-body model for the $f_{7/2}=7/2^-$ state
is consistent with that of the four-body
model of Ref.~\cite{2017Jones},\footnote{The energy of the 
$f_{7/2}=7/2^-$ state reported in Ref.~\cite{2017Jones}
is 4.77~MeV, and it is given with respect to the ground state. Thus, in
order to make a proper comparison with the results of
Ref.~\cite{2017Jones}, we have added to 4.77~MeV the energy of the ground state
$d_{3/2}=3/2^+$, resulting in 
$4.77~\text{MeV}+0.766~\text{MeV}=5.536~\text{MeV}$.} although
our pole width is much larger than that of Ref.~\cite{2017Jones}. 

\begin{figure} 
\begin{center}
   \includegraphics[angle=-90,width=0.65\columnwidth]{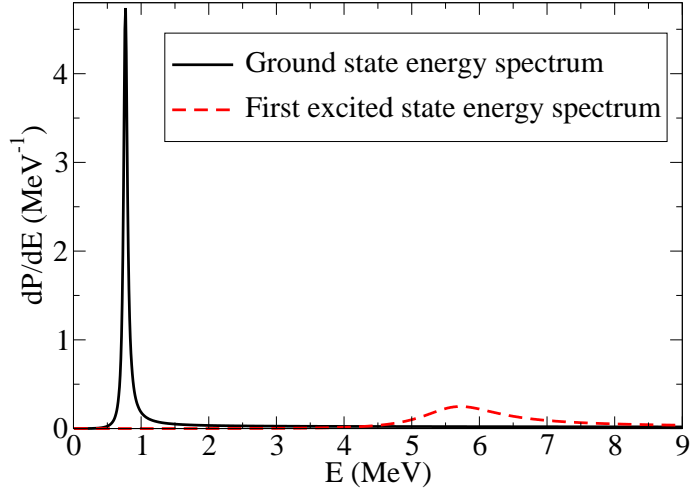}
    \caption[T]{
      (Color online) Decay energy spectrum of the ground state $3/2^+$ 
(solid, black line) and the first excited state $7/2^-$ (red, dashed line) of 
 $^{25}$O.}
     \label{fig.spectrum}
\end{center}
\end{figure}

Figure~\ref{fig.spectrum} shows the decay energy spectrum of
Eq.~(\ref{eq.des}) for the ground state $d_{3/2}=3/2^+$ and for the
first exited state $f_{7/2}=7/2^-$ of our two-body model. The spectrum
of the ground state is a narrow, sharp peak, whereas that of the first
excited state is less pronounced and much wider.

In Ref.~\cite{NPA15}, it was proposed that the decay widths can account for
the overall strength of the interaction between the resonance and the
continuum. In our case, $\overline{\Gamma}_{d_{3/2}}=0.133$~MeV, and
$\overline{\Gamma}_{f_{7/2}}=2.424$~MeV. Clearly, the coupling with the 
continuum is much stronger for the first excited state, which makes it less
sharp (i.e., less ``bound'') than the ground state. It was also proposed in 
Ref.~\cite{NPA15} that one could use the dimensionless decay constant
$\Gamma=\frac{\overline{\Gamma}}{\gr}$ as a measure of the coupling between the
resonance and the continuum: The larger $\Gamma$, the more ``bound'' the
resonance is, and hence the weaker the coupling to the continuum is. In our
case, $\Gamma _{d_{3/2}}=7.824$, and $\Gamma_{f_{7/2}}=6.956$. Thus, their
dimensionless decay 
constants also indicate that the ground state is more
``bound'' than the first excited state, as it should be.

The experimental decay energy spectrum of Ref.~\cite{2017Jones} was 
consistent with the inclusion of a first-excited 1/2$^+$ state of
$^{25}$O. The relative cross section of the ground
and first excited states was determined to be 
$\frac{\sigma_{3/2^+}}{\sigma_{1/2^+}} = 4$, which can be interpreted by
saying that producing the ground state is four times as likely as producing
the first excited state. Using the Gamow-state description of resonances,
it is possible to introduce a different way to quantify the relative
likelihood of production of two resonances. Since the decay width
$\overline{\Gamma}$ quantifies the overall strength of the coupling between
the resonance and the continuum, and since resonances with small decay
widths would be sharper than those with
larger decay widths, we can use the ratio of the decay widths of two resonances
as a measure of the relative strength of their decay energy spectra, i.e.,
as a measure of how likely one can observe the decay energy spectra of
a given resonance compared to that of another one. In our model,
$d_{3/2}=3/2^+$ is the ground state and $f_{7/2}=7/2^-$ is the first
excited state, and we have that
$\frac{\overline{\Gamma}_{f_{7/2}} }{\overline{\Gamma}_{d_{3/2}} }\sim 18$. This 
means that, according to our two-body model, it is much more likely to 
produce the ground state $d_{3/2}=3/2^+$ than the first-excited state 
$f_{7/2}=7/2^-$, as is already clear by visually inspecting their
spectra in Fig.~\ref{fig.spectrum}.

Theoretically, the width of the peaks of decay energy spectra are determined
by the pole widths $\gr$ rather than by the decay widths $\overline{\Gamma}$,
as can be seen in Fig.~\ref{fig.spectrum}. However, the theoretical pole 
widths are usually much smaller than the widths of the experimental decay energy
spectra. The reason is that the experimental resolution of the detector
is usually convoluted with the true decay energy spectrum, and such
convolution tends to broaden the spectrum. Thus, in order to compare
our theoretical formula with experiment, we would need to deconvolute
the experimental resolution from the true decay energy spectrum.

The resonant peaks in experimental decay energy spectra are usually fitted
with symmetric distributions, although there are examples of
asymmetric ones (see for example Refs.~\cite{2009Hoffman,2017LHCb}). However, 
as can be seen in Fig.~\ref{fig.spectrum}, the Gamow-state spectra are 
always slightly asymmetric. Such asymmetry is not part of any background, 
but arises 
from the energy dependence of the matrix element. Nevertheless, for sharp 
resonances the asymmetry is small. 

Experimental decay energy spectra are often fitted with a 
Breit-Wigner distribution that has an energy-dependent width, as was
done in Refs.~\cite{VANDEBROUCK17,HOFFMAN08,2009Hoffman,KOHLEY13,2017Jones}. It
seems therefore pertinent to discuss the similarities and differences between 
Eq.~(\ref{eq.des}) and the Breit-Wigner distribution with energy-dependent 
width. First, both approaches
yield quasi-Lorentzian peaks. However, in the case of Eq.~(\ref{eq.des}),
the Lorentzian is distorted by the matrix element of the interaction, 
whereas in the case of the Breit-Wigner distribution with energy-dependent width
the distortion is produced by the energy dependence of the width. Second, 
a fit of the experimental decay energy 
spectrum (after deconvolution of the experimental error) using
Eq.~(\ref{eq.des}) would yield quantitatively different resonant energies and 
widths than using the Breit-Wigner distribution with energy-dependent 
width. Third, and most importantly, an energy-dependent 
width $\Gamma (E)$ implies an
energy-dependent lifetime $\tau (E)=\frac{\hbar}{\Gamma(E)}$. Thus, 
although from a data-analysis point of view the Breit-Wigner
distributions with energy-dependent width may not seem very different 
from Eq.~(\ref{eq.des}), from a 
theoretical point of view they imply that resonances have different 
lifetimes for different energies. By contrast, when one describes a resonance 
by a Gamow state and its decay energy spectrum by Eq.~(\ref{eq.des}), the 
(mean) lifetime is given by $\tr=\frac{\hbar}{\gr}$, and it is an 
intrinsic property of the resonance that doesn't depend on its energy. Four, 
instead of the pole width $\gr$ being energy dependent,
in the Gamow-state approach what depends on the energy is the differential 
decay width $d\overline{\Gamma}(E)/dE$ of Eq.~(\ref{eq.ddw}), and such 
dependence takes into account that the resonance couples to the continuum with
different strengths at different energies, while at the same time
$\gr$, and therefore $\tr$, are energy independent.

In principle, it is possible to test experimentally whether Eq.~(\ref{eq.des})
or the Breit-Wigner distribution with energy-dependent width should be
used as the true theoretical decay energy distribution of resonances. One
could prepare the resonance at different energies, measure the mean lifetime
for each energy, and check whether the lifetime changes with energy (as in
Breit-Wigner distributions with energy-dependent width) or not 
(as in Eq.~(\ref{eq.des})).






\section{Conclusions} 
\label{sec.conclusions}

Decay energy spectra of radioactive nuclei are routinely measured, and 
theoretical nuclear models should be able to predict such spectra. Within 
the limitations of a simple two-body model of $^{25}$O, we have shown that 
the Gamow-state description of resonances is able to produce such spectra.
We have modeled the $^{25}$O nucleus as a 
valence neutron interacting with an $^{24}$O core, and applied
the formalism of Ref.~\cite{NPA15} to obtain the decay spectra of the
ground and first-excited states. The resulting spectra have a 
quasi-Lorentzian peak centered around the resonant energy, and 
are qualitatively
similar to those of simpler models~\cite{NPA17}, another example
of the universality of resonance phenomena. We have also seen that the
fits of experimental decay energy spectra that use a Breit-Wigner 
distribution with energy-dependent width are different, both quantitatively and 
phenomenologically, from fits that use the Gamow-state approach. In particular, 
the Breit-Wigner distributions with energy-dependent width imply an 
energy-dependent lifetime, whereas in the Gamow-state approach the lifetime 
is energy independent. We can, in principle, use the energy (non)dependence 
of the lifetime $\tr$ of a resonance to check whether the Gamow-state decay
spectrum or the Breit-Wigner distribution with an energy-dependent width should 
be used as the true theoretical decay energy spectrum.

There are several ways in which the results of the present paper can be
expanded. The most obvious one is the calculation of the decay energy
spectrum of an unstable nucleus using the Gamow Shell Model. In particular,
for $^{25}$O, most of the ingredients needed to calculate the decay energy 
spectrum have already been obtained~\cite{2017Fossez,2017Jones}, and it 
should be possible to obtain theoretical decay energy spectra that can be 
compared with experimental ones after deconvoluting the experimental 
error. Another way to extend the results of the present paper is by applying
the Gamow-state description of decay energy spectra to multi-channel
problems. There are methods to calculate the partial decay widths and 
branching fractions in a multi-channel system~\cite{2017Rakityansky}, but 
the calculation of the decay energy spectrum of a multi-channel potential 
is still awaiting.

\ack
The work of R.I.B.~was supported  by the National Council of Research
PIP-625 (CONICET, Argentina).

\appendix
\section{Appendix A: An almost delta-shell potential}
\setcounter{equation}{0}
\label{sec.appendix}

In this Appendix, we will explain in what sense the potential $V_a(r)$ of 
Eq.~(\ref{almostdelta}) can be considered the delta-shell
potential of strength $-V_0$ when $a$ tends to zero.

Let us consider the set of functions $K_n(r)$ defined as
\begin{equation}
     K_n(r)=-\frac{df_n(r)}{dr}=
   -\frac{d}{dr}\left(\frac{1}{1+e^{n(r-R)}}\right)=
        n \frac{e ^{ n(r-R)}}{\left( 1+e ^{n(r-R)} \right) ^2} \, , 
          \label{kns}
\end{equation}
where $n$ corresponds to $1/a$ in Eq.~(\ref{almostdelta}). Intuitively,
$-f_n(r)$ tends to a unit-step function located at $r=R$ when $n$ tends to
infinity (i.e., when $a$ tends to zero), so its derivative $K_n(r)$
should tend to the Dirac delta function $\delta (r-R)$.

In order for $K_n(r)$ to tend to the Dirac delta function, two conditions
are usually required:

\begin{itemize}
\item[({\it i})] When $n$ tends to infinity, the functions
$K_n(r)$ tend to zero everywhere, except at $r=R$, where they diverge.

\item[({\it ii})] The functions $K_n(r)$ are normalized to 1.
\end{itemize}

Condition ({\it i}) is clearly satisfied. In order to check 
condition~({\it ii}), let us calculate the following integral:
\begin{equation}
       \int_{r_{\rm min}}^{\infty} K_n(r) \, dr = -  
        \left[ f_n(r) \right] _{r_{\rm min}}^{\infty} 
        =  \frac{1}{1+e ^{n({r_{\rm min}}-R)}} \, .
\end{equation}
If $r_{\rm min} = -\infty$, then $\int_{-\infty}^{\infty} K_n(r) \, dr =1$,
condition~({\it ii}) is satisfied,
and $K_n(r)$ tends to $\delta (r-R)$ when $n$ tends to 
infinity. However, in our case $r_{\rm min}=0$,
$\int_0^{\infty} K_n(r) \, dr = \frac{1}{1+e ^{-nR}}\neq 1$, and therefore the
functions $K_n(r)$ do not satisfy condition~({\it ii}). Nevertheless,
as $n$ becomes large, $e ^{-nR}$ is very small,
and $\int_0^{\infty} K_n(r) \, dr =\frac{1}{1+e ^{-nR}}$, although not equal
to 1, is very close to 1, and we have that for the purposes of the present
paper the functions $K_n(r)$ become the delta function as $n$ tends to
infinity. This is why the potential $V_a(r)$ of 
Eq.~(\ref{almostdelta}) can be considered an almost delta-shell potential
of strength $-V_0$.

We can formalize the above discussion even further by using the theory of 
distributions~\cite{1966Schwartz}.

\vskip0.5cm

{\bf Definition.} {\it A set of functions $K_n(x)$ is a 
singular kernel that approximates the delta function at the origin if
\begin{equation}
        \lim_{n\to \infty}\int_{-\infty}^{\infty} K_n(x)\varphi (x) \, dx= \varphi (0)
       \label{definsk}
\end{equation}
for any test function $\varphi$.}

\vskip0.2cm

In bra-ket notation Eq.~(\ref{definsk}) can be written as 
\begin{equation}
   \lim_{n\to \infty}\langle K_n|\varphi\rangle = 
         \langle \delta|\varphi \rangle = \varphi (0) \, .
\end{equation}
Thus, definition~(\ref{definsk}) is the mathematical formalization of
calculating the Dirac delta function as the limit of functions.

\vskip0.5cm

{\bf Theorem.} {\it Let $K_n(x)$ be a sequence of locally integrable functions 
that satisfy the following conditions:

\begin{itemize}
   \item[(1)] There exists a positive $s$ such that $K_n(x)\geq 0$ when
$|x|<s$.
   \item[(2)] $K_n$ converges uniformly to zero in any set
$0<x_0\leq |x| \leq 1/x_0$ for any
$x_0>0$.
   \item[(3)] $\lim_{n\to \infty}\int_{|x|\leq x_0}K_n(x)\, dx =1$ for any
$x_0>0$.
\end{itemize}
Then the sequence $K_n(x)$ is a singular kernel that tends to the
delta function at the origin as $n \to \infty$.}

\vskip0.5cm

Condition~{\it (2)} means that as $n$ tends to infinity, the tails of
$K_n(x)$ become vanishingly small. Condition~{\it (3)} means that as $n$
tends to infinity, $K_n(x)$ is concentrated at the origin.

To apply the theorem to the functions in Eq.~(\ref{kns}), we need to use
$x=r-R$ and $x_0=r_0-R>0$. We will first apply the theorem to the case that
$r$ runs over the whole real line. Condition~{\it (1)} is 
clearly satisfied. When $x=r-R$ is negative, condition~{\it (2)} is also
clearly satisfied. When $x=r-R$ is positive, we have that
\begin{equation}
       K_n(x)= n \frac{e ^{ n(r-R)}}{\left( 1+e ^{n(r-R)} \right) ^2} <
       n\frac{1}{e ^{n(r-R)}} < n\frac{1}{e ^{n(r_0-R)}}
        \xrightarrow[n\to \infty]{\rm uniformly} 0
               \, .
\end{equation}
Hence, condition~{\it (2)} is also satisfied when $x>0$. Condition~{\it (3)}
is satisfied as well:
\begin{equation}
      \int_{-x_0}^{x_0} K_n(x)\, dx  = -
\left[f_n(x) \right]_{-x_0}^{x_0} =
\frac{-1}{1+e^{n(r_0-R)}}+\frac{1}{1+e^{-n(r_0-R)}} 
\xrightarrow[n\to \infty]{} 1
         \label{condition2}
\end{equation}

However, in our case $r$ is positive, $x$ is greater than
$-R$, and therefore the above theorem does not apply as stated. Nevertheless,
when $n$ becomes large, the tails of $K_n(r)$ are so small when $r<0$ that
conditions~{\it (2)} and~{\it (3)} are satisfied four our purposes, and we
have that the sequence $K_n(r)$ tends to $\delta (r-R)$ when $n$ tends to
infinity.

\bibliographystyle{elsarticle-num} 

\end{document}